# Experimental demonstration of Layered/Enhanced ACO-OFDM in short haul optical fiber transmission link

Binhuang Song, Chen Zhu, Bill Corcoran, *member, IEEE,* Qibing Wang, Leimeng Zhuang and Arthur J. Lowery, *Fellow, IEEE*

*Abstract*—Asymmetrically clipped optical orthogonal frequency division multiplexing (ACO-OFDM) is theoretically more power efficient but less spectrally efficient than DC-bias OFDM (DCO-OFDM), with less power allocating to the informationless bias component by only using odd index sub-carriers. Layered/Enhanced asymmetrically clipped optical orthogonal frequency division multiplexing (L/E-ACO-OFDM) has been proposed to increase the spectral efficiency of ACO-OFDM. In this letter, we experimentally demonstrate a 30-km single mode fiber transmission using L/E-ACO-OFDM at 4.375 Gbits/s. Using a Volterra filter based equalizer, 2-dB and 1.5-dB $Q$-factor improvements for L/E-ACO-OFDM comparing with DCO-OFDM can be obtained in back-to-back and 30-km fiber transmission respectively.

*Index Terms*—intensity modulation with direct detection (IM/DD), asymmetrically clipped orthogonal frequency division multiplexing (ACO-OFDM), optical communication.

## I. INTRODUCTION

SHORT-haul optical fiber communications usually adopt low-cost solution such as intensity modulation with direct detection. Several modulation technologies have been introduced to fit this real-value and non-negative link channel such as pulse amplitude modulation (PAM) [1], discrete multi-tone (DMT) or DCO-OFDM [2],[3], and ACO-OFDM [4],[5]. Compared with PAM, OFDM formats facilitate higher spectral efficiency and less sensitivity to dispersion and power fading if bit and power loading is implemented [6]. Among them, DCO-OFDM is able to guarantee non-negative signal by adding an adequate direct current bias. ACO-OFDM realizes it by only assigning data to odd sub-carriers and then clipping all negative values [5]. The clipping noise would fall into all the even sub-carriers which would not affect the detection performance. Unfortunately, this scheme reduces half of the data-carrying sub-carriers compared with DCO-OFDM. To regain the spectral efficiency, hybrid DCO/ACO-OFDM technology has been proposed in [7]. L/E-ACO-OFDM is an alternative way to increase spectral efficiency of ACO-OFDM by filling sub-carriers with different ACO-OFDM layers at transmitter and removing the clipping noise iteratively at the receiver. It has been proved to have a sensitivity advantage over DCO-OFDM in simulation [8],[9],[]. For practical implementations, however, a direct modulated laser (DML) must work with bias above the threshold current to provide modulated output [10],[11], which would reduce the power efficiency advantage of the intrinsically non-negative L/E-ACO-OFDM signal. In order to minimize the operating bias, a post-equalizer is needed to compensate the distortion induced by reducing the operating bias of the DML [Ref].

This letter provides the first experimental demonstration and comparison between L/E–ACO-OFDM and DCO-OFDM in a short-haul optical fiber transmission link. In this experiment, we implemented a truncated second-order Volterra filter for equalizing both the DCO-OFDM and L/E-ACO-OFDM signals. A noise cancellation algorithm [Ref] was also implemented for L/E-ACO-OFDM. The experimental result shows that L/E-ACO-OFDM can provide a 2-dB sensitivity improvement in back-to-back transmission and 1.5-dB advantage after 30-km fiber transmission over DCO-OFDM.

## II. L/E-ACO-OFDM & DCO-OFDM WAVEFORM GENERATION

Real-valued waveform of DCO-OFDM and L/E-ACO-OFDM systems can be generated by allocating complex data and its Hermitian conjugate value to positive and negative frequency sub-carriers before the inverse Discrete Fourier transform (IDFT) operation. To obtain the no-negative values, an extra DC bias is added for DCO-OFDM signals as shown in Fig. 1(b). However for L/E-ACO-OFDM, by systematically arranging each layer (or chord [12]) ACO-OFDM and then clipping all negative values to zero, the transmitted waveform becomes unipolar and all data can be recovered free of clipping noise at a cost of a slight spectral efficiency decrease. To be more specific, instead of loading data to all positive half of the IDFT size sub-carriers directly (in Fig. 1(a)), we load

This work is supported under the Australian Research Council's Laureate Fellowship (FL130100041) and CUDOS – ARC Centre of Excellence for Ultrahigh bandwidth Devices for Optical Systems (CE110001018).

Binhuang Song, Chen Zhu, Bill Corcoran, Qibing Wang, Leimeng Zhuang and Arthur J. Lowery are with the Department of Electrical and Computer Systems Engineering, Monash University, Melbourne, VIC 3800, Australia. Chen Zhu is currently with Bell Laboratories, Nokia, 600 Moutain Ave, Murray Hill, NJ 07974, USA. (e-mail: binhuang.song@monash.edu; chen.zhu@nokia.com; bill.corcoran@monash.edu;qibing.wang@monash.edu; leimeng.zhuang@monash.edu; arthur.lowery@ monash.edu).

1st-layer ACO-OFDM data to red sub-carriers (index 1+2$n$, $n$ = 0,1,2,3…), the 2nd-layer ACO-OFDM data to green sub-carriers (index 2+4$n$, $n$ = 0,1,2,3…), then the 3rd ACO-OFDM data to purple sub-carriers (index 4+8$n$, $n$ = 0,1,2,3…), as in Fig. 1(c). Then all negative values are set to zero before three layers' waveforms were superposed as in Fig. 1(d). To decode data in the receiver side, the 1st-layer will be first processed with a forward FT as it does not any suffer clipping noise from other layers, then its clipping noise (red triangles in Fig. 1(f)) is estimated and removed before processing the next layer. All layers' data can then be recovered by iteratively estimate and remove clipping noise to reveal next layer.

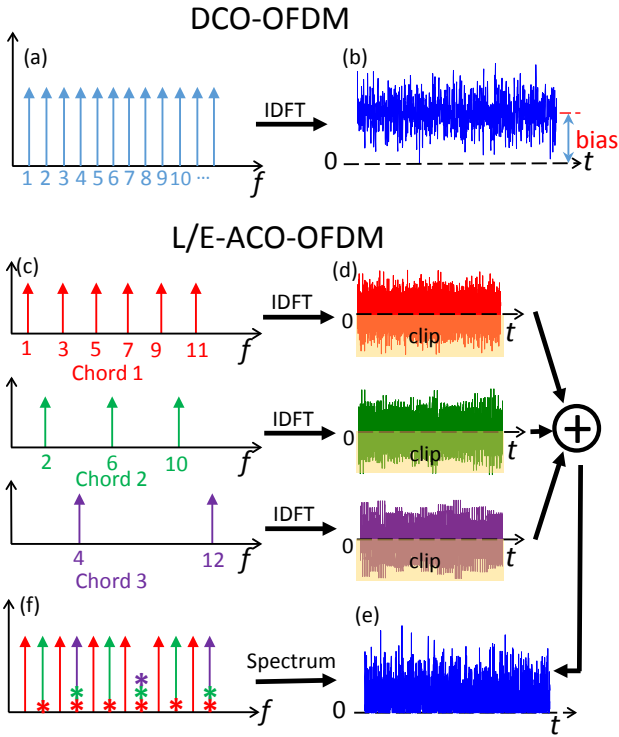

Fig. 1. (a) Frequency-domain and (b) time-domain conceptual diagrams for DCO-OFDM, (Arrows: sub-carriers). (c) Frequency-domain and (d) time-domain conceptual diagrams for each layer/chord in L/E-ACO-OFDM. (e) Time-domain and (f) frequency domain conceptual diagrams for total L/E-ACO-OFDM (Arrows: sub-carriers, star symbols: clipping noise).

## III. RECEIVER ALGORITHMS

L/E-ACO-OFDM can have a better receive sensitivity than that of DCO-OFDM because it does not need a large DC bias. Given a certain thermal noise in the system, a lower DC bias will give a higher signal-to-noise ratio (SNR) with a set received optical power (ROP). However, a DML typically needs a bias somewhat above its lasing threshold to work in its linear region, to avoid memory effects such as delayed turn-on after long off-periods and transient overshoots [10],[11].

Conventionally, a one-tap frequency domain equalizer is used for OFDM signals. However in this experiment, to compensate the above distortions, a Volterra filter time domain equalizer is implemented in the receiver.

### A. Truncated second-order Volterra filter

Volterra filters have been used to compensate chromatic dispersion-induced and chirp-induced distortions in optical communication systems and nonlinear distortion in high power amplifiers in wireless systems [3],[13],[14]. Because of the square-law detection in an IM/DD system, the implemented Volterra filter is limited to the second-order. The truncated Volterra filter can be expressed as:

$$y(k) = \sum_{l_1=0}^{L-1} w_1(l_1)x(k-l_1) + \sum_{l_1=0}^{L-1}\sum_{l_2=0}^{L-1} w_2(l_1,l_2)x(k-l_1)x(k-l_2),$$

where: $y$ is the output, $x$ is the input signal, $w_1$ and $w_2$ are weights for linear and second order terms. The Volterra filter also considers the multiplication terms of different samples. The above equation is a linear finite impulse response (FIR) filter with $L$ taps plus a second order FIR filter with $L(L+1)/2$ taps. Considering the memory length of the channel and computing complexity, we use $L$=10 for this experiment. A training sequence with a least-mean-square (LMS) algorithm is used to update the weights' values. This time domain equalizer can deal with arbitrary modulation formats as it only performs waveform correction.

### B. Noise cancellation algorithm for L/E-ACO-OFDM

For the L/E-ACO-OFDM signals, after time-domain Volterra equalization, an iterative noise cancellation algorithm is performed to further reduce the noise transfer between different layers (chords). The algorithm is similar to [15], which takes advantage of opposite-polarity property of ACO-OFDM waveforms. By doing a pairwise comparison of a1 and a1's (or a2 with a2') value (shown in Fig. 2) and setting the smaller one to zero, up half of the noise can be removed before processing the next layer (chord).

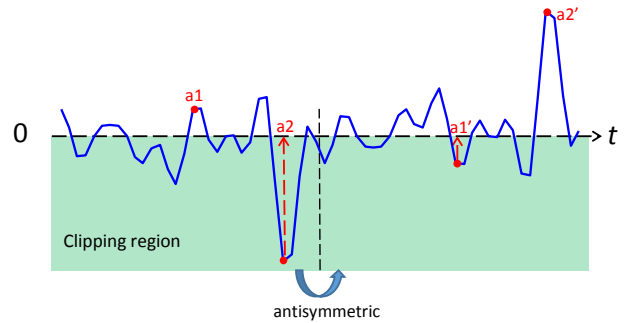

Fig. 2. Waveform of single layer/chord in L/E-ACO-OFDM. Blue trace: ACO-OFDM waveform before clipping. Any value in the green region will be clipped to zero. Before clipping, point a1 (a2) has same magnitude as a1' (a2') but with the opposite sign.

## IV. EXPERIMENTS

The experimental setup for 4.375 Gb/s QPSK transmission using DCO-OFDM and L/E-ACO-OFDM is shown in Fig. 3. An AA0701 DFB laser from Gooch & Housego was directly modulated in this experiment. The transmitted DCO-OFDM signal was generated in MATLAB with 256-point FFT, and has 63 sub-carriers carrying data and the 1st sub-carrier left for DC bias. For L/E-ACO-OFDM, we stack 3 layers (chords) with the same FFT size and oversampling rate, carrying 32, 16 and 8 sub-carriers for the 1st, 2nd and 3rd layers (chords). The L/E-ACO-OFDM's spectral efficiency is only 87.5% ((32+16+8)/64) of DCO-OFDM. In MATLAB, both the L/E-ACO-OFDM and DCO-OFDM signals were normalized so that their per-subcarrier powers are identical. Then the DCO-OFDM signal was clipped at 4 sigma before being fitted into a range from -0.5 V to 0.5 V in a Tektronix 7102 arbitrary waveform generator (AWG). The same scaling factor was applied for L/E-ACO-OFDM. Thus, because of the 50-Ω impedence of the laser and series resistor, both drive signals have 20-mA p-p amplitude. In order to achieve the same bit rate, the AWG sample rate was set to be 8.75 Gsample/s for DCO-OFDM and 10 Gsample/s for L/E-ACO-OFDM. The operating temperature of DML was controlled at 25 °C. Before the photodetector (PD), a power levelling attenuator controlled the total received optical power. The signal was sampled by a real-time digital oscilloscope (DSO) (Agilent DSO-X92804A) at 10 GSample/s.

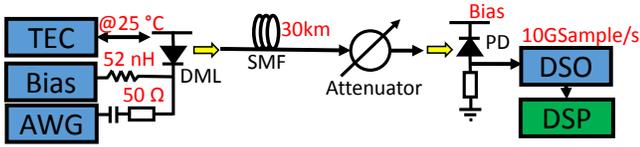

Fig. 3. Experimental setup. DSP: off-line digital single processing. AWG: arbitrary waveform generator. DML: direct modulated laser. TEC: temperature controller. SMF: single mode fiber. PD: photodetector. DSO: digital oscilloscope.

Fig. 4 shows the receiver DSP. We use either frequency-domain one-tap equalization or time-domain Volterra equalization (with noise cancellation for L/E-ACO-OFDM) for both systems. For each specific received optical power (ROP), we adjusted the DC bias to record the optimal performance (best $Q$-factor).

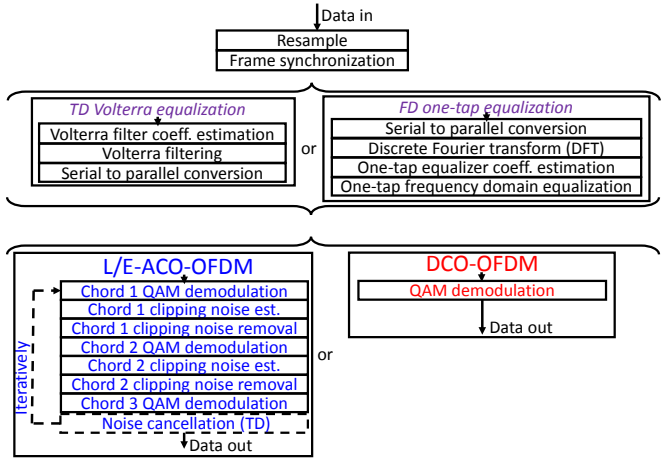

Fig. 4. Receiver DSP flowchart. Either time-domain Volterra equalization (left purple) and frequency-domain one-tap equalizing (right purple) are applied on L/E-ACO-OFDM (blue) and DCO-OFDM (red).

## V. RESULTS AND DISCUSSION

We first performed the back-to-back transmission experiment. At the first step, we set the received optical power to 0 dBm and found the optimal bias (20.891 mA for L/E-ACO-OFDM and 21.082 mA for DCO-OFDM). Then we measured the $Q$-factors at different ROPs while keeping these DC biases, shown in Fig. 5. In the low ROP region, where thermal noise (electrical noise) dominates, the signal distortions, a 2-dB drop in $Q$-factor can be observed per 1-dB reduction in ROP. At high ROPs, the $Q$-factor is limited by system imperfections such as quantization noise.

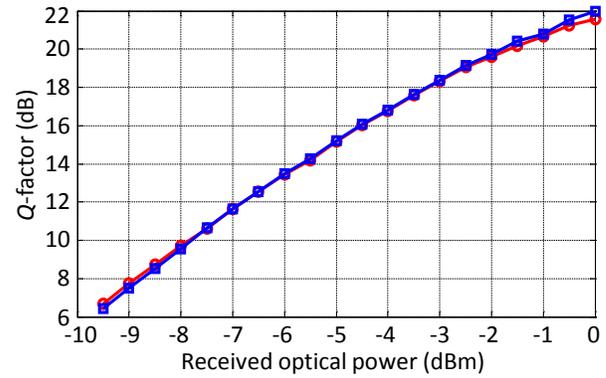

Fig. 5. $Q$-factor versus ROP for L/E-ACO-OFDM (blue squares) and DCO-OFDM (red circles).

Then, while tuning the DC bias, we measured the optimal $Q$-factors at different ROPs using time domain Voltera equalizing (with noise cancellation for L/E–ACO-OFDM) or one-tap frequency domain equalization methods. Fig. 6 shows a 2-dB $Q$-factor improvement for L/E-ACO-OFDM at -10

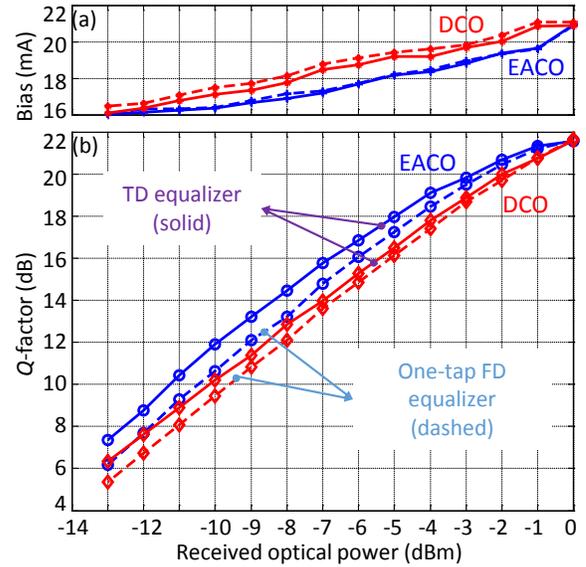

Fig. 6. (a) DC-bias for L/E-ACO-OFDM (blue lines) and DCO-OFDM (red lines). (b) $Q$-factor versus different ROPs for L/E-ACO-OFDM (blue line with circles) and DCO-OFDM (red line with diamonds). Solid line: using time domain equalizer. Dashed line: using frequency domain one-tap equalizer.

dBm ROP compared with DCO-OFDM.

Lastly, we compared the $Q$-factors for two OFDM formats in a 30-km single-mode fiber short-haul link. We did not include any cyclic prefix (CP) or bit/power loading algorithms. The measured $Q$-factors at local optimal biases are depicted in Fig. 7. It is clear that the second-order terms in the Volterra equalizer improve the signal quality in both OFDM formats. The L/E-ACO-OFDM has achieved a 1.5-dB $Q$-factor improvement at -10 dBm ROP over DCO-OFDM.

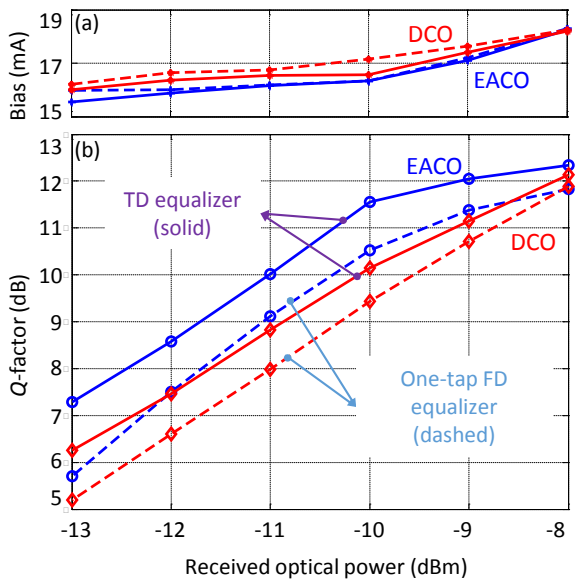

Fig. 7. (a) DC-bias for L/E-ACO-OFDM (blue lines) and DCO-OFDM (red lines). (b) $Q$-factor versus different ROP for L/E-ACO-OFDM (blue line with circles) and DCO-OFDM (red line with diamonds) after 30-km optical fiber transmission. Solid line: using time domain equalizer. Dashed line: using frequency domain one-tap equalizer.

In principle for QPSK formats, the L/E-ACO-OFDM should only have a 0.8-dB optical power advantage over optimally biased DCO-OFDM [5]. However, these theoretical bias levels are too low when driving a real laser in the experiment as the real DML has significant distortion near the lasing threshold (12.4 mA). From our results, L/E-ACO-OFDM requires less additional bias to reduce transient effects than DCO-OFDM. Thus, we see a slightly greater than expected benefit of using L/E-ACO-OFDM. As a result, a larger bias is needed for both DCO-OFDM and EACO-OFDM to improve the transient response of the laser but DC bias will reduce the signal-to-noise ratio for a given optical power. More $Q$-factor benefit is expected for a laser with lower threshold and less nonlinear distortion. A potential candidate is a vertical-cavity surfacing-emitting laser (VCSEL), as they normally have larger bandwidth and a low threshold [16],[17]. Higher-order Volterra filters could further improve the signal quality [18]. Other pre-compensation and pre-equalization algorithms for directly modulated laser transmission can be modified and integrated in this system [19] [20].

## VI. CONCLUSION

In this letter, we have demonstrated the first short-haul optical fiber transmission experiment using L/E–ACO-OFDM. A Volterra-filter based time-domain equalizer is implemented to suppress the distortion when operating DML at a low bias current. The experimental results shows that L/E–ACO-OFDM can have a 2-dB and 1.5-dB Q-factor advantage over DCO-OFDM in back-to-back transmission and 30-km optical fiber transmission respectively.